\documentclass[]{aastex62}

\defcitealias{finley2018dipquadoct}{FM18}
\usepackage{mhchem}

\usepackage{array,graphicx}
\usepackage{booktabs}
\usepackage{pifont}

\graphicspath{{./}{figures/}}

\received{--}
\revised{--}
\accepted{--}
\shorttitle{Solar Angular Momentum Loss With \textit{Wind}}
\shortauthors{Finley et al.}
\begin{document}

\title{Direct Detection of Solar Angular Momentum Loss with the \textit{Wind} Spacecraft}

\correspondingauthor{Adam J. Finley}
\email{*af472@exeter.ac.uk}

\author{Adam J. Finley}
\affil{University of Exeter,
              Exeter, Devon, EX4 4QL, UK}
\author{Amy L. Hewitt}
\affil{University of Exeter,
              Exeter, Devon, EX4 4QL, UK}
\author{Sean P. Matt}
\affil{University of Exeter,
              Exeter, Devon, EX4 4QL, UK}
\author{Mathew Owens}
\affil{University of Reading,
              Reading, Berkshire, RG6 6BB, UK}
\author{Rui F. Pinto}
\affil{IRAP, Universit\'e Toulouse III - Paul Sabatier,
            CNRS, CNES, Toulouse, France}
\author{Victor R\'eville}
\email{victor.reville@irap.omp.eu}
\affil{IRAP, Universit\'e Toulouse III - Paul Sabatier,
            CNRS, CNES, Toulouse, France}



\begin{abstract}
The rate at which the solar wind extracts angular momentum from the Sun has been predicted by theoretical models for many decades, and yet we lack a conclusive measurement from in-situ observations. In this letter we present a new estimate of the time-varying angular momentum flux in the equatorial solar wind, as observed by the \textit{Wind} spacecraft from 1994-2019. We separate the angular momentum flux into contributions from the protons, alpha particles, and magnetic stresses, showing that the mechanical flux in the protons is $\sim$3 times larger than the magnetic field stresses. We observe the tendency for the angular momentum flux of fast wind streams to be oppositely signed to the slow wind streams, as noted by previous authors. From the average total flux, we estimate the global angular momentum loss rate of the Sun to be $3.3\times10^{30}$erg, which lies within the range of various MHD wind models in the literature.  This angular momentum loss rate is a factor of $\sim$2 weaker than required for a Skumanich-like rotation period evolution ($\Omega_*\propto $ stellar age$^{-1/2}$), which should be considered in studies of the rotation period evolution of Sun-like stars.
\end{abstract}



\section{Introduction}

During the last $\sim4$ billion years, the Sun's rotation period is thought to have changed significantly due to the solar wind \citep{gallet2013improved, brown2014metastable, gallet2015improved, matt2015mass, johnstone2015stellar, amard2016rotating, blackman2016minimalist, sadeghi2017semi, see2018open, garraffo2018revolution, amard2019first}. This process, broadly referred to as \textit{wind braking}, appears to explain the observed rotation periods of many low-mass (i.e., $\leq 1.3M_{\odot}$), main-sequence stars \citep{skumanich1972time, soderblom1983rotational, barnes2003rotational, barnes2010simple, delorme2011stellar, van2013fast, bouvier2014angular}. Due to the interaction of the large-scale magnetic field on the outflowing plasma, this process is very efficient at removing angular momentum (AM), despite only a small fraction of a star's mass being lost to the stellar wind, during the main sequence \citep{weber1967angular, mestel1968magnetic, kawaler1988angular}.

Generally, the stellar magnetic field is thought of as providing a \textit{lever arm} for the wind, which many authors have attempted to quantify using results from magnetohydrodynamic (MHD) simulations \citep{matt2012magnetic, garraffo2015dependence, reville2015effect, finley2017dipquad, pantolmos2017magnetic, finley2018dipquadoct}. However, the AM loss rates from these MHD models have thus far been difficult to reconcile with the rates required by models of rotation-period evolution for low-mass stars \citep{finley2018effect,finley2019effect,see2019non}. Since many solar quantities are known to high precision (such as mass, radius, rotation rate and age), the Sun is often used to calibrate these \textit{rotation-period evolution} models. However, there are relatively few works that have attempted to model the current AM loss rate of the Sun \citep[e.g.,][]{alvarado2016simulating,reville2017global,finley2018effect,fionnagain2018solar,usmanov2018steady} and only a few studies that used in-situ measurements of the solar wind plasma and magnetic field \citep{lazarus1971observation, pizzo1983determination, marsch1984distribution, li1999magnetic}.  Consequently, the value of the solar AM loss rate remains uncertain, and the discrepancy between these two approaches remains in the literature.

The most direct, previous measurement of solar AM loss was performed using data from the two \textit{Helios} spacecraft by \cite{pizzo1983determination} and \cite{marsch1984distribution}. Despite requiring significant corrections to account for errors in spacecraft pointing, and using less than one year's worth of data, these authors were able to separate the individual contributions of the protons, alpha particles and magnetic field stresses. Interestingly, they showed that the alpha particles in the solar wind had an oppositely signed AM flux to the proton and magnetic components. Moreover, fast-slow stream-interactions appeared to transfer AM away from the fast component of the wind (causing the fast wind to often carry negative AM flux, like the alpha particles), which had also been noted by \cite{lazarus1971observation}. When compared, the contribution of the protons ($F_{AM,p}$) and magnetic field stresses ($F_{AM,B}$) were found on average to be comparable in strength ($F_{AM,p}/F_{AM,B} \sim 1$), although the AM flux in the protons was one of the most poorly determined components of the total flux. This result differs from previous work by \cite{lazarus1971observation} using the \textit{Mariner 5} spacecraft, who found the AM flux of the protons to dominate over the magnetic field stresses ($F_{AM,p}/F_{AM,B} \sim 4.3$). \cite{marsch1984distribution} showed that the ratio of AM flux in the particles and magnetic field stresses varies considerably with heliocentric distance and different solar wind conditions.

More recently, \cite{finley2018effect} combined observations of the solar wind (spanning $\sim20$ years) with a semi-analytic relation for the AM loss rate, derived from MHD simulations.  Theirs was a semi-indirect method, requiring in-situ measurements of only the mass flux and magnetic flux.  They found a global AM loss rate that varied in phase with the solar activity cycle, and had an average value of $2.3\times10^{30}$erg, compatible with the results from \cite{pizzo1983determination} and \cite{li1999magnetic} ($\sim3\times10^{30}$erg and $2.1\times10^{30}$erg respectively). By examining proxies of solar activity which span centuries and millennia into the Sun's past, \cite{finley2019solar} showed this value to be representative of the average over the last $\sim9000$ years. However, this value is lower than the  AM loss rate of $\sim6\times10^{30}$erg used in models that reproduce the rotational history of the Sun (and Sun-like stars) \citep{gallet2013improved, gallet2015improved, matt2015mass, finley2018effect, amard2019first}. Deviation from the rotational-evolution value has significant implications for our understanding of stellar rotation rates \citep{van2016weakened,garraffo2018revolution}, as well as for the technique of \textit{gyrochronology} \citep[e.g.][]{barnes2003rotational,metcalfe2019understanding}, in which stellar ages are derived from rotation rates.

In this letter, we provide a new direct measurement of the solar AM loss, which follows that of \cite{pizzo1983determination} and \cite{marsch1984distribution} but uses data from the \textit{Wind} spacecraft. These data span a period of $\sim$25 years and appear not to require the pointing corrections that were applied to the \textit{Helios} data.  This letter is organised as follows: in Section 2 we describe the data available from the \textit{Wind} spacecraft and calculate the time-varying mass flux and AM flux observed in the equatorial solar wind. Then in Section 3, we estimate the global AM loss rate and discuss the possible implications for the rotation period evolution of Sun-like stars.

\section{Observed Properties of the Solar Wind}

\subsection{Spacecraft Selection} \label{sec_selection}

The measurements required to accurately constrain the AM content in the solar wind particles are challenging to make \citep[see discussion in Section 3a of][]{pizzo1983determination}. Not only are the fluctuations in the AM flux comparable to the average value, but from an instrument standpoint, small errors in determining the wind velocity translate to large errors in the AM flux (because the radial wind speed is 2-3 orders of magnitude larger than the typical tangential speed of 1-10km/s at 1au). The latter problem appears to be the main reason why data from most spacecraft have not been used to measure AM \citep[see Figure 6 of][which shows data from the \textit{Ulysses} spacecraft; there is an approximately 1-year periodicty in the observations that is likely due to spacecraft pointing]{sauty2005nonradial}.  The magnetic field direction is generally more accurately determined because it is not as radial as the flow, and the instruments used are less sensitive to spacecraft pointing than the particle detectors (which get different exposures as the spacecraft pointing changes). Therefore, the magnetic stress component of the AM flux is typically better constrained.

While the {Advanced Composition Explorer (ACE)} spacecraft's non-radial solar wind speed measurements show the expected behaviours during periods of high variability \citep{owens2004b}, they appear to suffer from the same spacecraft-pointing-related issues as \textit{Ulysses} over longer time averages, in this case showing a strong $\sim$6-month periodicity. The \textit{Wind} and {Interplanetary Monitoring Platform 8 (IMP8)} spacecraft do not obviously show such features. Furthermore, during the period of overlap between \textit{Wind} and {IMP8}, there is good agreement in tangential wind speed, both in terms of the distributions and time series (linear regression of $r=0.81$ at the hourly time scale), suggesting limited instrumental effects.

In this work we focus on the high time cadence \textit{Wind} observations. {\textit{Wind} was designed to be a \textit{comprehensive solar wind laboratory for long-term solar wind measurements}, and has certainly stood the test of time; currently approaching its 25th year since launch (1st November 1994). During its mission lifetime the \textit{Wind} spacecraft completed multiple orbits of the Earth-Moon system, before relocating to a halo orbit about the $L_1$ Lagrangian point (on the Sun-Earth line) in May 2004. All the while collecting plasma and magnetic field measurements of the solar wind and Earth's magnetosphere with the Solar Wind Experiment \citep{ogilvie1995swe, kasper2006physics} and Magnetic Field Investigation instruments \citep{lepping1995wind}.}

\subsection{In-situ Measurements from the \textit{Wind} Spacecraft}
We analyse data recorded by the \textit{Wind} spacecraft\footnote{https://wind.nasa.gov/data.php - Data accessed in June 2019} from November 1994 to June 2019. Using data taken when the spacecraft was immersed in the solar wind, i.e.\ outside the Earth's magnetosphere. Additionally, we remove times when the spacecraft encountered Interplanetary Coronal Mass Ejections (ICMEs) using the catalogues\footnote{http://www.srl.caltech.edu/ACE/ASC/DATA/level3/icmetable2.htm - Data accessed in September 2019} of \cite{cane2003interplanetary} and \cite{richardson2010near} because ICMEs can produce large, non-radial, local flows that are not likely representative of global AM loss \citep{owens2004b}. For times not covered by the ICME catalogue (November 1994 - June 1996), we remove data with properties that are indicative of ICMEs, specifically data with a proton density greater than 70cm$^{-3}$ or field strengths greater than 30nT (a similar method was used by \citealp{cohen2011independency} on \textit{Ulysses} data).

Measurements of the solar wind magnetic field vector, proton density and velocity are available throughout the entire \textit{Wind} mission at $\sim2$ minute cadence. These parameters have a small number of entries flagged by the instrument team as containing unusable data, which we simply remove. Similarly, measurements of the alpha particle density and velocity are available, however the number of unusable data entries (where the proton and alpha particle populations cannot be deconvolved by the detector) is far greater. Therefore, when the alpha particles are flagged as unusable, we assume that the alpha particle density is $4\%$ of the proton density \citep[a representative value taken from][]{borrini1983helium} and that the alphas' velocities are identical to the protons'. We transform the vector quantities of velocity and magnetic field from GSE coordinates to RTN coordinates, where R points from the Sun to the spacecraft, T points perpendicular to the Sun's rotation axis in the direction of rotation, and N completes the right-handed triad \citep[further details are available in][]{franz2002heliospheric}.

For each quantity derived using \textit{Wind} data in this work, we calculate values at the smallest available cadence ($\sim2$ mins) and then average them over each Carrington Rotation (CR, $\sim27$-days) in our dataset. This helps to remove longitudinal variability caused by the rotation of features on the solar surface and smooths local fluctuations that occur on a range of shorter timescales. Finally, we require that each CR-average has more than $50\%$ of the data from that time period (after our cuts have been made). Otherwise, that CR is removed. In the top panel of Figure \ref{fluxes}, we plot the tangential wind speed of the protons and alpha particles as observed by \textit{Wind}.  For the tangential speeds shown in Figure \ref{fluxes}, we have weighted the CR averages by density, in order to reduce the obscuring effect of wind stream-interactions (see discussion in Section 3.3). Figure \ref{fluxes} shows typical tangential flow speeds of a few km/s, with variability that appears genuine and not to suffer from the errors present in data from other spacecraft (as discussed in Section  \ref{sec_selection}).

\subsection{Proton and Alpha Particle Properties}

\begin{figure*}
 \centering
  \includegraphics[trim=2cm 1cm 2cm 0.5cm, width=\textwidth]{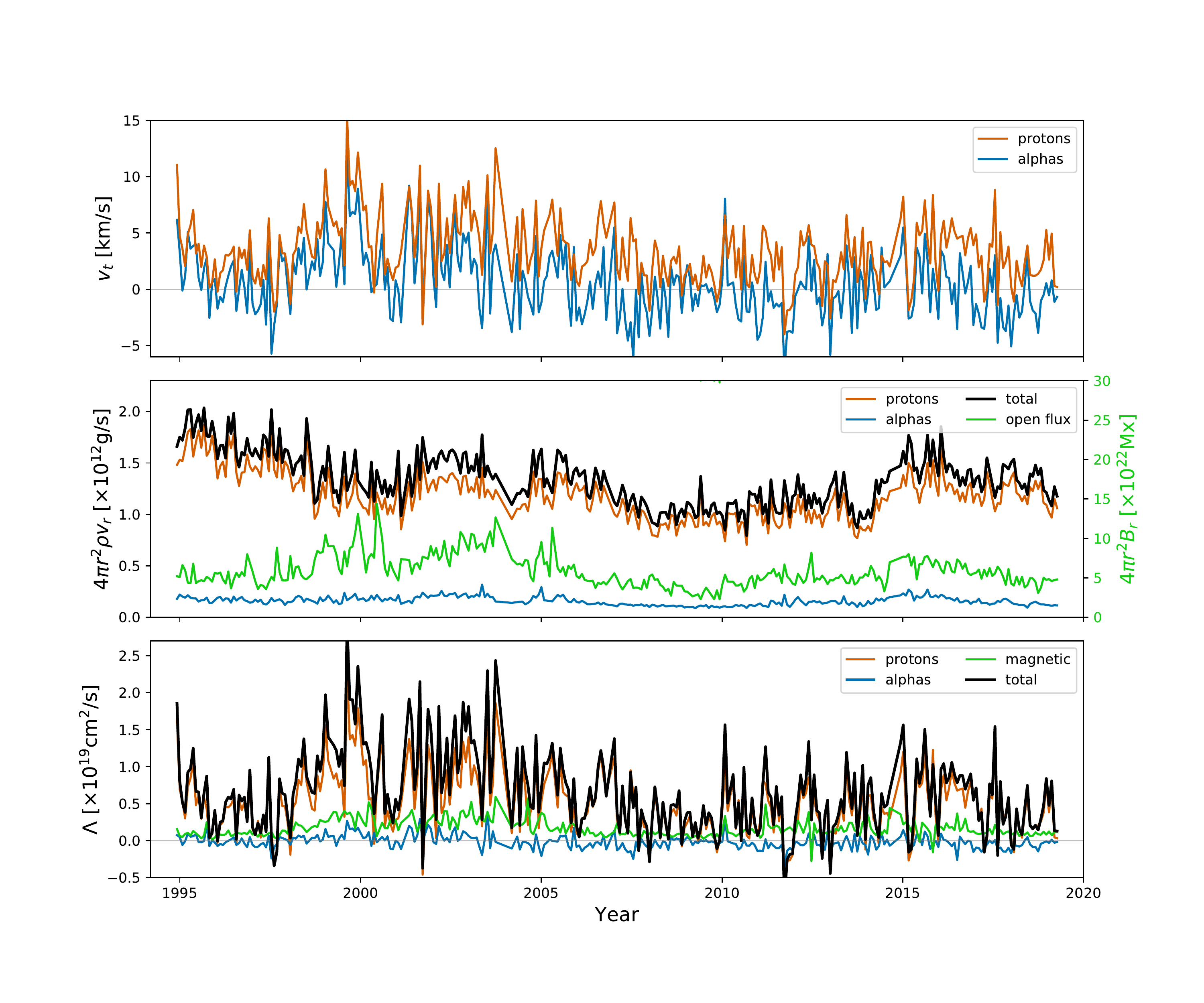}
   \caption{Top: CR-averages of the density-weighted, tangential speed of the protons and alpha particles in the solar wind versus time, plotted in orange and blue respectively. Middle: CR-averages of mass flux in the protons, alpha particles and their total (orange, blue and black lines), each multiplied by $4\pi r^2$, versus time. The prediction of equation (\ref{bopen}) for the open magnetic flux during the same time period is over-plotted using a green line, y-axis on the right {(see Section 3.1)}. Bottom: CR-averages of specific AM (defined as the AM flux per proton mass flux; density-weighted velocities are used here, see discussion) in the protons, alpha particles and magnetic field stresses (orange, blue and green lines) versus time. The total specific AM is plotted with a black line.}
   \label{fluxes}
\end{figure*}

The solar wind removes AM from the Sun at a rate proportional to the mass flux ($\rho v_r$) multiplied by the specific AM per unit mass ($\Lambda$). Using data from the \textit{Wind} spacecraft, we plot the mass flux in the protons, alpha particles and their total in the middle panel of Figure \ref{fluxes}. We multiply each by $4\pi r^2$ for an estimate of the global mass loss rate,
\begin{equation}
  \dot M \approx \langle 4\pi r^2(\rho_{p}v_{r,p} + \rho_{\alpha}v_{r,\alpha}) \rangle_{CR},
  \label{mdot}
\end{equation}
where the spacecraft's radial distance from the Sun is $r$, the radial wind speed is $v_r$, the solar wind density is $\rho$, the subscripts $p$ and $\alpha$ denote the proton and alpha particle components, and $\langle  \rangle_{CR}$ denotes an average over a ($\sim$27 day) CR. The total mass flux is dominated by the proton component of the wind and varies in a way that does not precisely correlate with the Sun's activity cycle \citep[see also][]{phillips1995ulysses,mccomas2000solar,finley2018effect,mishra2019mass}. By contrast, the alpha particle mass flux appears to be more strongly correlated with solar activity throughout the \textit{Wind} dataset {\citep[which is not surprising as the relative abundance of Helium in the equatorial solar wind is strongly correlated with solar activity, see][]{kasper2007solar}. }

%Though in general, the total mass flux is larger at solar maxima than minima, excluding the first 5 years of observations .

We define the specific AM as the AM flux divided by the proton mass flux (i.e., the specific AM per proton in the solar wind), which is given by,
\begin{equation}
    \Lambda = \bigg\langle r\sin\theta \bigg( v_{t,p} + v_{t,\alpha}\frac{\rho_{\alpha}v_{r,\alpha}}{\rho_p v_{r,p}} -\frac{B_{t}B_r}{4\pi\rho_p v_{r,p}} \bigg) \bigg\rangle_{CR},
    \label{lambda}
\end{equation}
where $\theta$ is the heliographic latitude of the spacecraft, $v_{t}$ is the tangential wind velocity, $B_{r}$ is the radial magnetic field strength and $B_{t}$ is the tangential magnetic field strength. {The first term in equation (\ref{lambda}) is the mechanical AM carried by the protons, the second term relates to the relative contribution of the alpha particles, and the final term describes the AM content of the magnetic field stresses.} Equation (\ref{lambda}) does not include the correction factor for the magnetic stresses which accounts for thermal pressure anisotropies, as it is expected to be negligible \citep[see][]{marsch1984helios}. In the bottom panel of Figure \ref{fluxes}, we plot the total specific AM along with the individual proton, alpha particle and magnetic field components. We use density-weighted tangential velocities, as in the top panel of Figure \ref{fluxes}, to reduce the effect of wind stream-interactions (see discussion in Section 3.3). Figure \ref{fluxes} shows the protons to dominate the specific AM of the solar wind, with the magnetic field stresses and alpha particles carrying much less specific AM (per proton).

%We find the specific AM flux to fluctuate strongly with solar activity unlike the mass flux, which remains relatively constant (see the middle panel of Figure \ref{fluxes}).

\subsection{Angular Momentum Flux Detection}

\begin{figure*}
 \centering
  \includegraphics[trim=2cm 0.4cm 2cm 0.5cm, width=\textwidth]{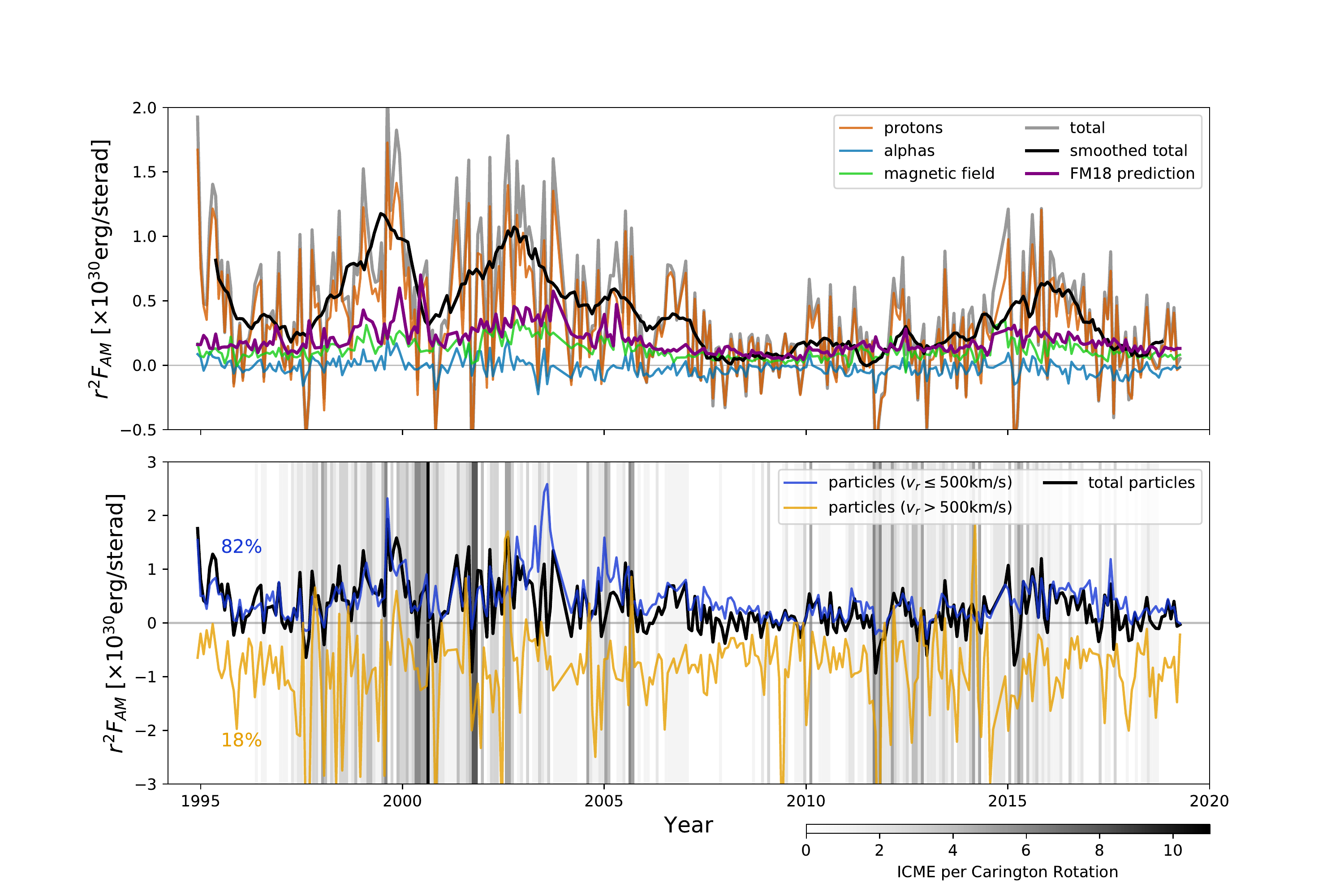}
   \caption{Top: CR-averages of AM flux multiplied by radial distance squared, versus time. The proton, alpha particle and magnetic components are shown with orange, blue and green lines respectively. The total of these is indicated with a grey line. A 13 CR moving average is shown with a thick black line. The prediction of the AM loss rate prescription of \cite{finley2018effect} (from equations  (\ref{open_torque}) and (\ref{FM18eq})) is shown with a purple line. Bottom: Similar plot as above, now only showing the particle component (protons plus alphas). We plot the average AM flux for particles with a radial speed less than, and greater than, 500km/s, in blue and yellow respectively. On average the \textit{Wind} spacecraft encountered the slower wind $82\%$ of the time. The number of near-Earth ICMEs per CR is shown with a color gradient in the background (following the colorbar at bottom-right).}
   \label{AMdetection}
\end{figure*}

The total AM flux in the protons, alpha particles and magnetic field stresses is given by multiplying the specific AM by the proton mass flux,
\begin{equation}
   F_{AM} = \langle \rho_p v_{r,p} \Lambda \rangle_{CR} = \bigg\langle r\sin\theta\bigg( \rho_{p}v_{r,p}v_{t,p} + \rho_{\alpha}v_{r,\alpha}v_{t,\alpha} - \frac{B_{t}B_{r}}{4\pi} \bigg) \bigg\rangle_{CR}.
   \label{AMflux}
\end{equation}
We plot the AM fluxes (multiplied by radial distance squared) in the protons, alphas, magnetic field and their total in the top panel of Figure \ref{AMdetection}. There is a large scatter/variability in the AM flux, despite averaging over whole CRs. The variability is mainly due to the varying specific AM (i.e., in the tangential wind speed), rather than changes in the mass flux (see Figure \ref{fluxes}), and which is likely affected by local fluctuations in the solar wind, caused by transients \citep{roberts1987origin, tokumaru2012long}. The solid black line in Figure \ref{AMdetection} shows a 13-CR (i.e., $\sim$1-year) moving average on the total AM flux, which more clearly describes the longer-term variability of the AM flux.  Our dataset contains sunspot cycles 23 and 24 (left and right halves of the Figures, respectively), which have notable differences in their AM fluxes.  Generally, during times of increased solar activity the specific AM of the protons and magnetic field stresses increase together, such that $F_{AM,p}/F_{AM,B}$ does not vary with solar activity. We find cycle 24, which is currently in its declining phase, has a much lower average AM flux than cycle 23 ($\sim40\%$ of cycle 23).

The average value for the AM flux, and that of each constituent, is listed and compared to previous estimates in Table \ref{averages}.  The \textit{Wind} total is primarily composed of the proton and magnetic field components, with the alpha particles contributing a small and mostly negative AM flux contribution.  In comparison with the work of \cite{pizzo1983determination} and \cite{marsch1984distribution}, the \textit{Wind} data show a much stronger AM flux in the protons and a large reduction (in amplitude) to the AM flux carried by the alpha particles. These differences could be related to long-term change in the solar wind.  For example, the solar wind appears denser in the last decade compared to the \textit{Helios} era \citep[see][]{mccomas2013weakest}. Or alternatively, due to the exchange of momentum between protons and alphas as the wind propagates into the heliosphere \citep[for which there is some evidence in][]{sanchez2016very}.

The AM flux in the magnetic field stress in the \textit{Wind} data is similar to that determined by \cite{pizzo1983determination} and \cite{marsch1984distribution} but is smaller than determined by \cite{lazarus1971observation}. Interestingly, the dominant contribution to the \textit{Wind}-measured AM flux comes from the protons, with the magnetic field of secondary importance. In simplified MHD simulations of the solar wind (such as those of \citealp{finley2017dipquad}), the ratio $F_{AM,p}/F_{AM,B}$ depends on parameters such that the larger the Alfv\'en radius ($R_A$) the larger the contribution of the magnetic field. The average ratio measured by \textit{Wind} is $F_{AM,p}/F_{AM,B}=2.6$, which is significantly different than the ratio of $\sim$1 found by  \cite{pizzo1983determination}. \cite{marsch1984distribution} showed that \textit{Helios} data from smaller heliocentric distances gives larger ratios, which might account for the difference.  The proton-dominated regime shown by the \textit{Wind} data is consistent with MHD simulations that have cylindrically-averaged $R_A$ smaller than $15R_{\odot}$.

%Given the equatorial AM flux we find in this work, we use the relation $R_A = (\Lambda/\Omega_{\odot})^{1/2}$ from \cite{weber1967angular} to infer an equatorial Alfv\'en radius of $30-40R_{\odot}$, which appears to contradict this.

\section{Discussion}

Using data from the \textit{Wind} spacecraft, we have evaluated the flux of AM in the equatorial solar wind. In this section, we estimate the global AM loss rate of the Sun and compare with an MHD model and rotational evolution models. Additionally, we discuss the effect of ICMEs and interacting wind streams on our dataset.
\subsection{Comparison to Theory}
To show our result in the context of current theoretical predictions, we compare to the AM loss rate of \cite{finley2018effect}, which was derived using MHD simulations. In their work, the AM loss rate is given by,
\begin{eqnarray}
    \dot J_{FM18} = (2.3\times10^{30}[\text{erg}])\bigg(\frac{\dot M}{1.1\times 10^{12} [\text{g/s}]}\bigg)^{0.26}
    \times \bigg(\frac{\phi_{open}}{8.0\times 10^{22}[\text{Mx}]}\bigg)^{1.48},
    \label{open_torque}
\end{eqnarray}
where the AM loss rate of the Sun is parameterised in terms of the mass loss rate, $\dot M$, and the open magnetic flux, $\phi_{open}$. The open magnetic flux in the solar wind is estimated by,
\begin{equation}
    \phi_{open} = \langle 4\pi r^2 |B_r|_{1hr}\rangle_{CR},
    \label{bopen}
\end{equation}
where the average value of the radial magnetic field is assumed to be representative of the global open magnetic flux in the solar wind. This assumption has been discussed by many previous authors \citep{wang1995solar,lockwood2004open,pinto2017multiple} and has observational support \citep{smith1995ulysses,owens2008estimating}. Using equation (\ref{bopen}) we plot the open magnetic flux using data from the \textit{Wind} spacecraft in the middle panel of Figure \ref{fluxes} with a solid green line.

Using equation (\ref{open_torque}) we calculate the predicted AM loss rate of the solar wind, where the mass loss rate and open magnetic flux (equations (\ref{mdot}) and (\ref{bopen})) are calculated using data from the \textit{Wind} spacecraft. We then relate the AM loss rate and AM flux using,
\begin{equation}
  \dot J = \oint_A {\bf F_{AM}}\cdot d{\bf A} = F_{AM,eq} \int_0^{2\pi}\int_0^{\pi} r^2(\sin\theta)^3 d\theta d\phi,
  \label{jdot}
\end{equation}
where A represents a closed surface in the heliosphere (we adopt a sphere of radius $r$), $\phi$ is heliographic longitude, and $F_{AM,eq}$ is the AM flux in the solar equatorial plane, assumed to be equivalent to that measured by CR-averages of data taken in the ecliptic. As the AM flux in the solar wind is expected to vary with latitude, we have assumed a physically motivated functional form\footnote{If the wind is spherically symmetric, the latitude dependence can be understood by considering the proton term in equation (\ref{AMflux}), where a geometric factor of $\sin\theta$ appears at the start of the equation to compute the cylindrical radius. Another geometric factor of $\sin\theta$ appears from the approximation of solid body rotation (i.e., $v_{t}\propto\sin\theta$). }, ${\bf F_{AM}}(\theta) \approx  F_{AM,eq}(\sin\theta)^2{\bf \hat{r}}$. By rearranging equation (\ref{jdot}) we produce a relation for the equatorial AM flux,
\begin{equation}
  F_{AM,eq}= \frac{\dot J}{2\pi r^2\int(\sin\theta)^3 d\theta} \approx \frac{\dot J}{2.7\pi r^2}.
  \label{FM18eq}
\end{equation}
The AM flux from equation (\ref{FM18eq}), using the AM loss rate from equation (\ref{open_torque}), is plotted with a solid purple line in the top panel of Figure \ref{AMdetection}. Strikingly, this result matches well during solar minimum wind conditions. However it consistently under-estimates the AM flux during solar maxima. The \cite{finley2018effect} AM loss rates were derived from simulations with only one wind acceleration profile, but differing wind acceleration profiles have been shown to affect the predicted AM loss rates \citep{pantolmos2017magnetic}. Therefore changes in the balance of fast and slow wind in the heliosphere are not taken into account by this model. It is known that the proportion of slow wind changes significantly from solar minimum to maximum, whilst the ecliptic remains essentially dominated by the slow wind the whole time (\textit{Wind} encountered slow wind streams, with $v_r<500$km/s, $82\%$ of the time). Importantly, this implies that the \textit{Wind} observations may be more representative of global conditions at solar maximum than solar minimum.  Uncertainties in our assumed latitudinal distribution of AM flux prevent us from producing a more conclusive estimate of the global AM loss rate. For us to better constrain this, there is a need for simultaneous observations at higher latitude (e.g., combined measurements with both the \textit{Wind} spacecraft and the upcoming Solar Orbiter) but at present, the current approach is the best we can do without introducing further uncertainty.

\begin{table}
\caption{Mean of the CR-averaged Solar Angular Momentum Fluxes}
\label{averages}
\center
\setlength{\tabcolsep}{2pt}
  \begin{tabular}{cccc}
      \hline\hline
Component &$\langle r^2 F_{AM}\rangle$ 	& Source 	&	Citation	\\
 &[$\times 10^{30}$erg/ster]	&  	&		\\	\hline
Protons & $0.29^{}_{}$ & \textit{Wind} & This work \\
& 0.17 & \textit{Helios} & \cite{pizzo1983determination} \\
& $\sim1$ & \textit{Mariner 5} & \cite{lazarus1971observation} \\ \hline
Alpha Particles & $-0.02^{}_{}$ & \textit{Wind} & This work \\
& $-0.13$ & \textit{Helios} & \cite{pizzo1983determination} \\ \hline
Magnetic Field & $0.12^{}_{}$ & \textit{Wind} & This work \\
& 0.15 & \textit{Helios} & \cite{pizzo1983determination} \\
& 0.23 & \textit{Mariner 5} & \cite{lazarus1971observation} \\ \hline
Total & $0.39^{}_{}$ & \textit{Wind} & This work \\
& 0.20 & \textit{Helios} & \cite{pizzo1983determination} \\
& $0.26^{}_{}$ & Theory & This work, equations  (\ref{open_torque}) and (\ref{FM18eq}) \\ \hline
  \end{tabular}
\end{table}

  \subsection{Implications for the Rotation Evolution of Sun-like Stars}

    Rearranging equation (\ref{FM18eq}) produces an estimate of the global AM loss rate based on the average AM flux detected by the \textit{Wind} spacecraft, $\dot J_{\textit{Wind}} = 2.7\pi\langle r^2  F_{AM}\rangle=3.3\times10^{30}$erg. This AM loss rate is approximately half that required by the empirical Skumanich relationship, where rotation period evolves proportional to the square root of stellar age \citep{skumanich1972time}.  Specifically, for the Sun's rotation to follow the Skumanich relationship, the present-day AM loss rate must be $\approx 6.2\times10^{30}$erg \citep{finley2018effect}. The torque-averaged Alfv\'en radius, $R_A =\sqrt{\dot J/(\dot M \Omega)}$, implied by the \textit{Wind} result is $R_A \approx 15R_{\odot}$, in contrast to $R_A \approx 20R_{\odot}$ using the AM loss rate required for Skumanich-like rotation. We note the value of $R_A$ from \textit{Wind} is in better agreement with MHD simulations that reproduce the observed ratio of $F_{AM,p}/F_{AM,B}$ (see Section 2.4).

    The (unknown) systematic uncertainties in our result could be large enough to resolve this discrepancy.  However, taken at face value, and assuming the Sun is not special, our result could be evidence that stars deviate significantly from the Skumanich relationship at around the solar age \citep[or Rossby number, for example, as suggested by][]{van2016weakened}.  Alternatively, our result could mean that the present-day solar wind is in some kind of ``low state,'' such that the AM loss rate averaged over timescales of $\gg 25$ years is significantly larger (see \citealp{finley2019solar} and \citealp{finley2018effect} for a discussion and other caveats).

    %This solar AM loss rate implies a spin-down time of $\sim18$ gigayears, assuming a moment of inertia for the Sun of $6.90\times10^{53}$g cm$^2$ from the models of \cite{baraffe2015new}.

\subsection{Coronal Mass Ejections and Fast-slow Stream-interactions}

Detecting the AM flux is complicated by the myriad of transients and fluctuations in the solar wind. With sufficient spatial averaging of the heliosphere (or sufficient temporal averaging at a fixed location), the contribution of transients to the AM flux is likely to be small. However, with the available observations, large transient structures can bias estimates of the AM flux. In this work we have attempted to remove times when ICMEs interacted with the \textit{Wind} spacecraft. We show the number of near-Earth ICMEs per CR as a colour gradient in the bottom panel of Figure \ref{AMdetection}, which is well correlated with solar activity. The plasma properties of ICMEs are often very different to the ambient wind, typically having stronger magnetic fields and increased mass fluxes. Surprisingly, if we include these events in our calculation, the computed equatorial AM flux decreases by 4$\%$.  Although we have been careful to remove such events, ICME catalogues are not perfect, and therefore errors due to ICMEs are more likely to be introduced in times of high solar activity, or times where no ICME catalogues are available (i.e. November 1994 - June 1996).

Additionally, as noted by previous authors \citep{lazarus1971observation,pizzo1983determination,marsch1984distribution}, our results contains evidence for fast-slow wind interactions. The net effect of these interactions is expected to be zero, given sufficient averaging. We plot the average AM flux in the solar wind particles with radial wind speeds greater and less than 500km/s separately in the lower panel of Figure \ref{AMdetection}. The slower component of the wind, when compared with the total particle AM flux plotted in black, is shown to carry the bulk of the AM flux in the particles. The faster component is shown to have a mostly small or negative AM flux. However this component does not strongly contribute to the total AM flux during each CR because of the small fraction (on average $18\%$) of the time \textit{Wind} encountered this flow, but also because fast wind streams tend to carry smaller mass flux, further reducing their contribution to the total AM flux.

This dichotomy between faster and slower wind streams occurs because of interactions within the solar wind as it propagates into interplanetary space. When fast and slow wind streams ``collide'', the slow wind undergoes an acceleration in the direction of corotation and the fast component is deflected oppositely \citep[see Figure 1 in][]{pizzo1978three}. Though most of this acceleration occurs in the radial direction, some is directed tangentially. The impact this has on our fluxes is far more pronounced in the faster component because it is typically less dense than the slower component. This effect makes the tenuous AM flux signal harder to distinguish when simply looking at the raw tangential wind speeds, and has been shown to become increasingly important with increasing heliocentric distances \citep[see Figure 2 in][]{marsch1984distribution}. Since \textit{Wind} data are taken at $\sim1$au, and in the equatorial plane (where stream-interactions are expected to be more pronounced), we chose to present the tangential wind speeds and specific AM in Section 2 weighted by density. Doing so produces values that are more representative of their contribution to the AM flux.

\section{Conclusion}

In this letter we have attempted to measure the current AM loss rate of the Sun, using data from the \textit{Wind} spacecraft to directly evaluate the equatorial AM flux in the solar wind. Our findings are summarised as follows:
\begin{enumerate}
  \item The strongest contribution to the AM flux at $\sim1$au comes from the protons, which carry on average $\sim75\%$ of the total flux. Our result is similar to that of \cite{lazarus1971observation} using the \textit{Mariner 5} spacecraft ($\sim80\%$), and some of the measurements from the \textit{Helios} spacecrafts at smaller heliocentric distances of $\sim0.3$au \citep{marsch1984distribution}.
  \item Both the alpha particles and fast ($v_r>500$km/s) wind components contribute a negative source of AM flux (at $\sim1$au), most likely resulting from dynamical processes in the solar wind. We find the alpha particles carrying a much smaller AM flux than \cite{pizzo1983determination} found in the \textit{Helios} data.
  \item The average equatorial AM flux is $0.39\times10^{30}$erg/sterad, which lies within the predictions of various current theoretical works.  The equatorial AM flux varies with solar cycle and during solar maxima is observed to be significantly larger than the predictions of \cite{finley2018effect}.
  \item We estimate the global AM loss rate of the Sun to be $3.3\times10^{30}$erg, which is a factor of $\sim2$ smaller than is expected from a Skumanich-like rotation period evolution of a Sun-like star. It is difficult to conclude whether this discrepancy indicates a weakened braking (e.g., as inferred by \citealp{van2016weakened}), or is due to differences in the latitudinal distribution of AM flux from our assumed profile, or is perhaps indicative of long-time variability in the AM loss rate of the Sun \citep[see][]{finley2019solar}.
\end{enumerate}
We are hopeful that missions such as Parker Solar Probe \citep{fox2016solar} and Solar Orbiter \citep{mueller2013solar} will begin to provide valuable data towards addressing the AM loss rate of the Sun. Specifically, Parker Solar Probe is sampling the solar wind at distances where stream-interactions are expected to be weaker (or not formed yet), and the signal to noise should be enhanced.

\acknowledgments
We thank the instrument teams who contributed to the \textit{Wind} spacecraft, and the NASA/GSFC's Space Physics Data Facility for providing this data.
AJF, ALH and SPM acknowledge funding from the European Research Council (ERC) under the European Union’s Horizon 2020 research and innovation programme (grant agreement No 682393 AWESoMeStars).
MO is funded by Science and Technology Facilities Council (STFC) grant numbers ST/M000885/1 and ST/R000921/1.
RFP acknowledges support from the French space agency (Centre National des Etudes
624 Spatiales; CNES; https://cnes.fr/fr).
VR acknowledges funding by the ERC SLOW{\_}\,SOURCE project (SLOW{\_}\,SOURCE - DLV-819189).
Figures in this work are produced using the python package matplotlib \citep{hunter2007matplotlib}.




\bibliographystyle{yahapj}
\bibliography{Adam}



\end{document}